\documentclass[12pt]{article}
\usepackage{a4}
\usepackage{amsmath}
\usepackage{amsfonts}
\usepackage{amsmath}
\usepackage{latexsym}
\usepackage{amsfonts}
\usepackage{graphics}
\usepackage{graphicx}
\usepackage{epic}
\usepackage{eepic}

\begin{document}
\renewcommand{\thefootnote}{\fnsymbol{footnote}}
\thispagestyle{empty}

\noindent hep-th/0506185   \hfill  June   2005 \\

\noindent \vskip3.3cm
\begin{center}

{\Large\bf The quantum one loop trace anomaly  of the higher spin
conformal conserved currents in  the bulk of $AdS_{4}$}
\bigskip\bigskip\bigskip

{\large Ruben Manvelyan\footnote{On leave from Yerevan Physics
Institute} and Werner R\"uhl}
\medskip

{\small\it Department of Physics\\ Erwin Schr\"odinger Stra\ss e \\
Technical University of Kaiserslautern, Postfach 3049}\\
{\small\it 67653
Kaiserslautern, Germany}\\
\medskip
{\small\tt manvel,ruehl@physik.uni-kl.de}
\end{center}

\bigskip 
\begin{center}
{\sc Abstract}\end{center} An analysis of the structure and
singularities of the one loop two point function of the higher spin
traceless and conserved currents constructed from the single scalar
field in $AdS$ space is presented. The detailed renormalization
procedure is constructed  and the quantum violation of the traceless
Ward identity is investigated. The connection with the one loop
effective action for higher spin gauge fields is discussed.

\newpage

\section{Introduction}
The increasing interest in the complicated problem of quantization
and interaction of the higher spin gauge theories in $AdS$ space
\cite{Frons, Vasiliev} is connected  with the $AdS_{4}/CFT_{3}$
correspondence of the critical $O(N)$ sigma model and four
dimensional higher spin gauge theory in anti de Sitter space (HS(4))
proposed in \cite{Klebanov}. This special case of general $AdS/CFT$
correspondence can be used for direct reconstruction of the unknown
bulk interaction from the well developed boundary theory \cite{R,T}.
This unique case is interesting also in view of the properties of
the renormalization group flow from the free field \emph{unstable}
point of the boundary $O(N)$ vector model with the stable critical
interacting conformal point in the large -$N$ limit by the
deformation with the double trace marginal operator. This flow
should correspond to the quantum behaviour  on
 the bulk side of the same higher spin theory
($HS(4)$) and different boundary conditions for the quantized scalar
field \cite{WittKleb}. Note that in the second and nontrivial
conformal point of the $d=3$ sigma model all higher spin currents
except the energy-momentum tensor (spin two) are conserved only in
the large $N$ limit and their divergence is of first order in
$\frac{1}{N}$. On the bulk side this must correspond to a certain
mass generation mechanism on the one loop level (again order of
$\frac{1}{N}$) of the interacting $HS(4)$ gauge theory. This mass
generation mechanism was considered in our previous articles
\cite{Ruehl, RM3}. In this article we want to consider the bulk
mirror of the instability of the free field conformal point of the
boundary theory. The main idea is the following: Because any
interaction pulls the free theory out of the conformal point with an
infinite number of the higher even spin conserved traceless currents
(corresponding to the traceless higher spin gauge fields on the
bulk) any one loop self energy graph constructed using any possible
(self)interaction of the corresponding gauge higher spin field on
the bulk should violate tracelessness of the latter. Here we will
investigate only the simplest possible minimal gauge field times
current interaction of the HS field with the bulk scalar field
constructed in  \cite{MR} and consider the behaviour of the short
distance singularities in the coordinate space of the corresponding
one loop two point function of the conserved currents responsible
for the renormalization of the propagator of the HS gauge fields. We
prove that correct regularization and renormalization leads to the
violation of the traceless Ward identities when we maintain the
quantum level conservation Ward identities (quantum gauge
invariance). We call this phenomenon \emph{Higher spin trace
anomaly} due to the analogy with the conformal trace anomaly of the
energy momentum tensor which is the spin equal two case of our
general spin consideration.

\section{Conserved current in AdS}
Using our notation of the previous papers \cite{RM3,MR,RM2} we
consider the minimal interaction of the conformal higher spin field
with the conserved traceless current constructed from the
conformally coupled scalar in $AdS_{4}$ space\footnote{We will use
Euclidian $AdS_{d+1}$ with conformal flat metric, curvature and
covariant derivatives satisfying
\begin{eqnarray}
&&ds^{2}=g_{\mu \nu }(z)dz^{\mu }dz^{\nu
}=\frac{L^{2}}{(z^{0})^{2}}\delta _{\mu \nu }dz^{\mu }dz^{\nu
},\quad \sqrt{g}=\frac{L^{d+1}}{(z^{0})^{d+1}}\;,
\notag  \label{ads} \\
&&\left[ \nabla _{\mu },\,\nabla _{\nu }\right] V_{\lambda }^{\rho
}=R_{\mu \nu \lambda }^{\quad \,\,\sigma }V_{\sigma }^{\rho }-R_{\mu
\nu \sigma
}^{\quad \,\,\rho }V_{\lambda }^{\sigma }\;,  \notag \\
&&R_{\mu \nu \lambda }^{\quad \,\,\rho
}=-\frac{1}{(z^{0})^{2}}\left( \delta _{\mu \lambda }\delta _{\nu
}^{\rho }-\delta _{\nu \lambda }\delta _{\mu }^{\rho }\right)
=-\frac{1}{L^{2}}\left( g_{\mu \lambda }(z)\delta _{\nu
}^{\rho }-g_{\nu \lambda }(z)\delta _{\mu }^{\rho }\right) \;,  \notag \\
&&R_{\mu \nu }=-\frac{d}{(z^{0})^{2}}\delta _{\mu \nu }=-\frac{d}{L^{2}}%
g_{\mu \nu }(z)\quad ,\quad R=-\frac{d(d+1)}{L^{2}}\;.  \notag
\end{eqnarray}%
For shortening the notation and calculation we contract all rank
$\ell$ symmetric tensors with the $\ell $-fold tensor product of a
vector $a^{\mu }$.}.

\setlength{\unitlength}{0.252mm}
\begin{picture}(854,108)(0,-168)
        \allinethickness{0.252mm}\dashline{6}(18,-114)(102,-114) 
        \put(0,-138){$h^{(\ell)}$} 
        \put(144,-72){$\sigma$} 
        \allinethickness{0.252mm}\special{sh 0.3}\put(102,-114){\ellipse{4}{4}} 
        \put(138,-168){$\sigma$} 
        \allinethickness{0.252mm}\path(102,-114)(162,-78) 
        \allinethickness{0.252mm}\path(102,-114)(162,-144) 
        \put(198,-120){$=$} 
        \put(264,-120){$S^{(\ell)conf}_{int}=\frac{1}{\ell}\int d^{4}x\sqrt{g}h^{(\ell)\mu_{1}\dots\mu_{\ell}}J^{(\ell)}_{\mu_{1}\dots\mu_{\ell}}.$} 
\end{picture}

Here $h^{(\ell)}$ is the spin $\ell$ gauge field and
$J^{(\ell)}(a;z)=J^{(\ell)}_{\mu_{1}\dots\mu_{\ell}}a^{\mu_{1}}\dots
a^{\mu_{\ell}}$ the conserved traceless current constructed from the
conformally coupled scalar $\sigma(z)$ \cite{MR} in $D=d+1$
dimensional $AdS$ space
\begin{eqnarray}
  &&J^{(\ell)}(z;a) = \frac{1}{2}\sum^{\ell}_{p=0}A_{p}\left(a\nabla\right)^{\ell
  -p}\sigma(z)\left(a\nabla\right)^{p}\sigma(z) \nonumber\\
  &&+ \frac{a^{2}}{2}\sum^{\ell -1}_{p=1}B_{p}\left(a\nabla\right)^{\ell
  -p-1}\nabla_{\mu}\sigma(z)\left(a\nabla\right)^{p-1}\nabla^{\mu}\sigma(z)\label{ansatz}\\
  &&+\frac{a^{2}}{2L^{2}}\sum^{\ell -1}_{p=1}C_{p}\left(a\nabla\right)^{\ell
  -p-1}\sigma(z)\left(a\nabla\right)^{p-1}\sigma(z) + O(a^{4}) +
  O(\frac{1}{L^{4}})\;,\nonumber
\end{eqnarray}
where $A_{p}=A_{\ell-p}, B_{p}=B_{\ell-p}, C_{p}=C_{\ell-p}$ and
$A_{0}=1$. The tracelessness condition
\begin{equation}\label{trace}
    \Box_{a}J^{(\ell)}(z;a)=\frac{\partial^{2}}
    {\partial a_{\mu}\partial a^{\mu}}J^{(\ell)}(z;a)=0 \; ,
\end{equation}
 fixes relations between
$B_{p}, C_{p}$ and $A_{p}$ in the following way \cite{Leonhardt}
\begin{eqnarray}
  &&B_{p}=-\frac{p(\ell -p)}{(D+2\ell-4)}A_{p} ,\label{flatT} \\
  &&C_{p}=\frac{-1}{2(D+2\ell -4)}\left[s_{t}(p+1,\ell,D)A_{p+1}
  +s_{t}(\ell -p+1,\ell, D)A_{p-1}\right], \quad\label{L2T}\\
  &&s_{t}(p,\ell,D)=\frac{1}{4}p(p-1)D(D-2) +
  \frac{1}{3}p(p-1)(p-2)(\ell +2D-5) \;.
\end{eqnarray}
The unknown $A_{p}$ can be fixed using the conservation condition
for the current
\begin{equation}\label{curcons}
\nabla \cdot
\partial_{a}J^{(\ell)}(z;a)=\nabla^{\mu}\frac{\partial}{\partial
a^{\mu}}J^{(\ell)}(z;a)=0 .
\end{equation}

This leads to a recursion relation with the same solution for the
$A_{p}$ coefficients \cite{Anselmi} as in the flat $D=d+1$
dimensional case
\begin{equation}\label{so1}
    A_{p}=(-1)^{p}\frac{\binom{\ell}{p}\binom{\ell +D-4}{p+\frac{D}{2}-2}}
    {\binom{\ell +D-4}{\frac{D}{2}-2}} \;.
\end{equation}
For the important  case $D=4$  this formula  simplifies to
\begin{equation}\label{4d}
 A_{p}=(-1)^{p}\binom{\ell}{p}^{2} \;.
\end{equation}
The result of our previous consideration \cite{Leonhardt} was the
following: the curvature corrections do not change flat space
tracelessness  and conservation conditions between leading
coefficients and therefore the solution (\ref{so1}) remains valid.
So we can concentrate in the future only on the  first part of our
currents described by the first set of coefficients (\ref{4d})($A$
terms) in four dimensional $AdS$ space
\begin{equation}\label{anz}
J^{(\ell)}(z;a) =
\frac{1}{2}\sum^{\ell}_{p=0}A_{p}\left(a\nabla\right)^{\ell
  -p}\sigma(z)\left(a\nabla\right)^{p}\sigma(z)+\textnormal{B\&C terms}
\end{equation}
knowing for sure that all trace ($B$-) and curvature ($C$-) terms
are not essential for quantum dynamics considered here and could be
restored from kinematical considerations and commutation relations
of covariant derivatives in $AdS$ space. In the center of our
interest here we will put the following self-energy one loop diagram
constructed from the $A$ term of our interaction current in
coordinate space

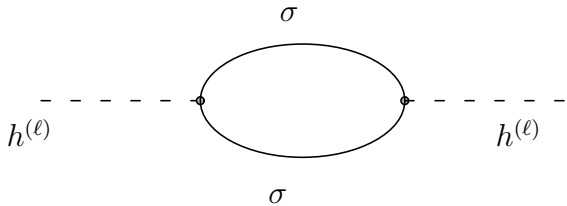
\setlength{\unitlength}{0.252mm}
\quad\quad\quad\quad\quad\quad\begin{picture}(311,108)(126,-168)
        \allinethickness{0.252mm}\dashline{6}(144,-114)(228,-114) 
        \put(126,-138){$h^{(\ell)}$} 
        \put(384,-138){$h^{(\ell)}$} 
        \put(270,-72){$\sigma$} 
        \allinethickness{0.252mm}\special{sh 0.3}\put(228,-114){\ellipse{4}{4}} 
        \allinethickness{0.252mm}\special{sh 0.3}\put(336,-114){\ellipse{4}{4}} 
        \allinethickness{0.252mm}\dashline{6}(336,-114)(420,-114) 
        \allinethickness{0.252mm}\put(282,-114){\ellipse{108}{60}} 
        \put(264,-168){$\sigma$} 
\end{picture}

This diagram is connected to the two point function of the currents
in the standard way
\begin{equation}\label{tpf}
    \int_{z_{1}} \int_{z_{2}}
    h^{(\ell)}(z_{1};a)\overleftarrow{\partial_{a}}\cdot
     \overrightarrow{\partial_{a}}< J^{(\ell)}(z_{1};a) J^{(\ell)}(z_{2};c) >
     \overleftarrow{\partial_{c}}\cdot\overrightarrow{\partial_{c}}h^{(\ell)}(z_{2};c).
\end{equation}
The quantum one loop behaviour, singularity and renormalization of
this two point function will be explored in the next sections.
\section{Loop function and Ward identity}

For the calculation of the one loop two point function
\begin{equation}\label{lf}
  \Pi^{(\ell)}(z_{1};a|z_{2};c):=  < J^{(\ell)}(z_{1};a) J^{(\ell)}(z_{2};c) >
\end{equation}
we have to insert corresponding $A$ terms of currents (\ref{ansatz})
in (\ref{lf}) and apply just Wick's theorem. The propagator of the
scalar field in $AdS_{4}$ quantized with a boundary condition
corresponding to the free conformal point of the boundary $O(N)$
model is\footnote{From now on we put $L=1$.}
\begin{equation}\label{prs}
<\sigma(z_{1})
\sigma(z_{2})>=\frac{1}{8\pi^{2}}\left(\frac{1}{\zeta-1}+\frac{1}{\zeta+1}\right)
,
\end{equation}
where
\begin{equation}\label{zeta}
    \zeta=\frac{(z_{1}^{0})^{2}+(z_{2}^{0})^{2}+
 (\vec{z}_{1}-\vec{z}_{2})^{2}}{2z^{0}_{1}z^{0}_{2}}
\end{equation}
and $\zeta-1$ is the invariant geodesic distance in $AdS$. Due to
this property our correlation function depends only on the $AdS$
invariant geodesic distance and it's derivatives exactly as in the
case of the higher spin propagator described in our previous article
\cite{RM2}. The general rule for working with such objects is
analyzed in detail in the same article. The main point is the
following. The tensorial structure of any two point function in
$AdS$ space can be described using a general  basis  of the
independent bitensors \cite{AllenJ}, \cite{AllenT}, \cite{Turyn},
\cite{LMR1}
     \begin{eqnarray}
       && I_{1}(a,c):=(a\partial)_{1}(c\partial)_{2}\zeta(z_{1},z_{2}) , \\
       && I_{2}(a,c):=(a\partial)_{1}\zeta(z_{1},z_{2})(c\partial)_{2}\zeta(z_{1},z_{2}),\\
       && I_{3}(a,c):=a^{2}_{1}I^{2}_{2c}+c^{2}_{2}I^{2}_{1a} , \\
       && I_{4}:=a^{2}_{1}c^{2}_{2} ,\\
       && I_{1a}:=(a\partial)_{1}\zeta(z_{1},z_{2})\quad ,
       \quad I_{2c}:=(c\partial)_{2}\zeta(z_{1},z_{2}) , \\
       &&(a\partial)_{1}=a^{\mu}\frac{\partial}{\partial
       z_{1}^{\mu}} ,\quad (c\partial)_{2}=c^{\mu}\frac{\partial}{\partial
       z_{2}^{\mu}} ,\\&& a^{2}_{1}=g_{\mu\nu}(z_{1})a^{\mu}a^{\nu} ,
       \quad c^{2}_{2}=g_{\mu\nu}(z_{2})c^{\mu} c^{\nu} .
     \end{eqnarray}
In this case this basis should appear automatically after
contractions of scalars and action of the vertex  derivatives. In
general we have to get an expansion with all four basis elements
\begin{equation}\label{K}
    \Pi^{\ell}(z_{1};a|z_{2};c)=\Psi^{\ell}[F]+\sum_{n,m;\, 0<2(n+m)<\ell}I_{3}^{n}I_{4}^{m}
    \Psi^{\ell-2(n+m)}[G^{(n,m)}] .
\end{equation}
Here we introduce a special map from the set
$\{F_{k}(\zeta)\}^{\ell}_{k=0}$ of the $\ell+1$ functions of $\zeta$
to the space of $\ell\times \ell$ bitensors
\begin{equation}\label{Psy}
    \Psi^{\ell}[F]=\sum^{\ell}_{k=0}I^{\ell-k}_{1}(a,c)I^{k}_{2}(a,c)F_{k}(\zeta)
    .
\end{equation}
But because for the analysis of short distance singularities ($\zeta
\rightarrow 1$) only the $A$ terms of currents are important for us,
we will restrict our consideration on the first part of (\ref{K})
connected with the $I_{1}, I_{2}$ bitensors  calling all monomials
corresponding to $I_{3}$ and $I_{4}$ in the above sum and the
corresponding sets of functions
$\{G^{(n,m)}_{k}\}_{k=0}^{\ell-2(n+m)}$ the "trace terms"
\begin{eqnarray}
  \Pi^{\ell}(z_{1};a|z_{2};c)=\Psi^{\ell}[F]+ \textnormal{trace terms} .\label{sets}
\end{eqnarray}
So all our calculations will be with exception of $O(a^{2})$ and
$O(c^{2})$ terms. These trace terms in principle can be analyzed
using the computer program \cite{LMR2}. All important relations for
the calculations to be performed below can be found in \cite{RM2}
and listed in Appendix A of this article. In the main text we will
only present the so called general multigradient map for the scalar
function of $\zeta$
\begin{eqnarray}
  && (c\cdot\nabla)_{2}^{q}(a\cdot\nabla)_{1}^{p}F(\zeta)
  =\sum^{q}_{n=0}\frac{p!\binom{q}{n}}{(p-q+n)!}
  F^{(p+q)}(\zeta)I_{1}^{q-n}I_{1a}^{p-q+n}I_{2c}^{n}+\textnormal{traces} ,\quad\quad \label{multigrad}\\
  &&F^{(k)}(\zeta):=\frac{\partial^{k}}{\partial\zeta^{k}}F(\zeta)\quad
  ,\quad p\geq q .
\end{eqnarray}

Now we are ready to calculate the correlation function (\ref{lf}).
Substituting (\ref{anz}) in (\ref{lf}) and using (\ref{prs}),
(\ref{4d}) and (\ref{multigrad})  after some manipulations we obtain
the following formula for our two point function
\begin{eqnarray}
  && \Pi^{\ell}(z_{1};a|z_{2};c)=\frac{1}{2^{7}\pi^{4}}\sum^{\ell}_{k=0}
  I_{1}^{\ell-k}I_{2}^{k}R_{k}^{\ell}(\zeta)+\textnormal{traces} ,\label{r1} \\
  &&R_{k}^{\ell}(\zeta)=\sum^{\ell+k}_{r=0}Q^{\ell}_{k,r}
  \left\{\Phi^{sing}_{k}(\zeta)+\Phi^{mixed}_{r,k}(\zeta)+\Phi^{reg}_{k}(\zeta)\right\} ,
  \label{sep}\\
  &&Q^{\ell}_{k,r}=\frac{(-1)^{\ell+k}(\ell!)^{2}(\ell+k)!}{(k!)^{2}(\ell-k)!}\sum^{r}_{p,q=r-k}
  \frac{(-1)^{p+q}\binom{\ell}{p}\binom{\ell}{q}\binom{k}{r-p}\binom{k}{r-q}\binom{\ell-k}
  {p+q-r}}{\binom{l+k}{r}} ,\label{Q}\\
  &&\Phi^{sing}_{k}(\zeta)=\frac{1}{(\zeta-1)^{\ell+k+2}} ,\label{sing}\\
  &&\Phi^{reg}_{k}(\zeta)=\frac{1}{(\zeta+1)^{\ell+k+2}} ,\label{reg}\\
  &&\Phi^{mixed}_{k,r}(\zeta)=\frac{2}{(\zeta-1)^{r+1}(\zeta+1)^{\ell+k-r+1}}\quad;\quad
  \left(Q^{\ell}_{k,r}=Q^{\ell}_{k,\ell+k-r}\right), \label{mixed}
\end{eqnarray}
where in (\ref{sep}) we separated the singular, regular and mixed
parts in view of their short distance behaviour at $\zeta\rightarrow
1$.

For the investigation of the short distance singularities at
$\zeta\rightarrow 1$ we have to take into account the following:
\begin{itemize}
  \item $I_{1}(a,c;\zeta)\rightarrow \frac{(a\cdot
  c)}{(z_{1}^{0})^{2}}$ if $\zeta\rightarrow 1$ .
  \item $I_{2}(a,c;\zeta)\rightarrow 0$ but
  $\frac{I_{2}(a,c;\zeta)}{\zeta-1}$ is finite when $\zeta\rightarrow
  1$ .
  \item Singularities start from $(\zeta-1)^{-2}$ because in $D=4$
  $\sqrt{g}d^{4}z$ behaves as $(\zeta^{2}-1)d\zeta$  (will be shown in Appendix
  A).
\end{itemize}
The next important point is the singular part of the mixed terms
coming from (\ref{mixed}). These terms are expanded as
\begin{eqnarray}
  && \Psi^{(\ell)mixed}_{sing}:=\Psi^{(\ell)}[F^{(\ell)}_{sing}] ,\label{ms}\\
  && F^{(\ell)}_{k,sing}=\frac{1}{2^{7}\pi^{4}}\sum^{\ell+1}_{m=2}a^{(\ell)}_{k,m}(\zeta-1)^{-m-k} ,\label{ms1}\\
  && a^{(\ell)}_{k,m}=2^{-\ell-1+m}\sum^{\ell+1-m}_{r=0}(-1)^{\ell+1-m-r}
  \binom{\ell+1-m}{r}Q^{\ell}_{k,r} .
\end{eqnarray}

In Appendix B we prove that the bitensor (\ref{ms}) formed by the
singular mixed part can be expressed as a bigradient of a spin
$\ell-1$ bitensor
\begin{equation}\label{mix1}
\Psi^{(\ell)mixed}_{sing}[F^{(\ell)}_{sing}]=(a\cdot\nabla_{1})(c\cdot\nabla_{2})
\Psi^{(\ell-1)}[G] ,
\end{equation}
formed by the set of functions $G_{k}^{(\ell-1)}(\zeta)$. Moreover
this procedure can be continued recursively in $\ell$ if we separate
the singular part of $\Psi^{(\ell-1)}[G]$ and express the latter as
a gradient term. The final formula is
\begin{equation}\label{multigrad1}
\Psi^{(\ell)mixed}_{sing}=\sum^{\ell}_{n=1}[(a\cdot\nabla_{1})(c\cdot\nabla_{2})]^{n}
\sum^{\ell-n}_{k=0}I_{1}^{\ell-n-k}I_{2}^{k}b^{(\ell-n)}_{k,n}(\zeta-1)^{-k-1}
.
\end{equation}
Both (\ref{mix1}) and (\ref{multigrad1}) are derived in Appendix B.
 By (\ref{multigrad1}) the singular part of $\Psi^{(\ell)mixed}$ is
 expressed as a sum of powers of bigradients applied to regular
 (integrable) functions. The whole expression (\ref{multigrad1})
 is therefore a well-defined distribution and does not need any
 regularization if we apply extracted derivatives on
 nonsingular external higher spin gauge fields in the effective
 action. This allows us to concentrate on the main singular part
 of the correlation function
(\ref{r1})
\begin{eqnarray}
  &&  \Pi^{\ell}(a,c;\zeta)=\frac{1}{2^{7}\pi^{4}}\sum^{\ell}_{k=0}
 \frac{ I_{1}^{\ell-k}I_{2}^{k}}{(\zeta-1)^{\ell+k+2}}\sum^{\ell+k}_{r=0}
 Q^{\ell}_{k,r} ,\label{main}
\end{eqnarray}
which can not be presented in such a form (\ref{mix1}),
(\ref{multigrad1}).

 The crucial point here is the possibility to sum
the coefficients $Q^{\ell}_{k,r}$ because the main singularity
(\ref{sing}) does not depend on the index $r$. Indeed we can observe
the following important identity
\begin{equation}\label{id}
    \sum^{\ell+k}_{r=0}Q^{\ell}_{k,r}=(-1)^{\ell+k}(2\ell)!\binom{\ell}{k}
    ,
\end{equation}
or explicitly
\begin{equation}\label{ident}
\frac{(\ell+k)!}{(k!)^{2}(\ell-k)!}\sum^{\ell+k}_{r=0}\sum^{r}_{p,q=r-k}
  \frac{(-1)^{p+q}\binom{\ell}{p}\binom{\ell}{q}\binom{k}{r-p}\binom{k}{r-q}\binom{\ell-k}
  {p+q-r}}{\binom{l+k}{r}}=\binom{2\ell}{\ell}\binom{\ell}{k} .
\end{equation}
Unfortunately we have no analytic proof of identity (\ref{ident})
but we are absolutely sure that it is right because we have checked
this strange identity for many possible numbers $\ell$ and $k$ with
a computer program (Mathematica 5). Thus using (\ref{id}) we can
immediately sum (\ref{main}) and obtain the following nice relation
\begin{equation}\label{singcor}
\Pi^{\ell}(a,c;\zeta)=\frac{(-1)^{\ell}(2\ell)!}{2^{7}\pi^{4}
}\left(I_{1}-\frac{I_{2}}{\zeta-1}\right)^{\ell}\frac{1}{(\zeta-1)^{\ell+2}}
.
\end{equation}

The beauty  of the expression (\ref{singcor}) is the following: This
main singular part of our loop function is satisfying the naive Ward
identities following from the conservation and tracelessness
conditions of our currents (\ref{trace}),(\ref{curcons}) directly
without contribution of the corresponding partner trace terms
described by the expansion in the other two bitensors $I_{3},I_{4}$
\begin{equation}\label{Ward}
\Box_{a}\Pi^{\ell}(a,c;\zeta)=(\nabla\cdot\partial_{a})\Pi^{\ell}(a,c;\zeta)=0
.
\end{equation}
Here $\Pi^{\ell}$ is considered as analytic function on $\zeta$. In
the next section we will introduce a correct regularization and
renormalization of (\ref{singcor}) and investigate the quantum
violation of the tracelessness condition.

\section{Extraction of singularities and renormalization}

Now we introduce the correct regularization for extracting the
singularities from (\ref{singcor}). For convenience we can use
instead of $\zeta$ as equivalent invariant variable - the chordal
distance
\begin{equation}\label{u}
    u=\zeta-1=\frac{(z_{1}-z_{2})^{2}}{2z_{1}^{0}z_{2}^{0}} .
\end{equation}
For this variable the singularity is located at $u\rightarrow 0$.
Then expanding again (\ref{singcor}) in the form
\begin{eqnarray}\label{dist}
    &&\Pi^{\ell}(a,c;u)=\frac{(-1)^{\ell}(2\ell)!}{2^{3}(4\pi)^{4}
}\sum^{\ell}_{k=0}
 I_{1}^{\ell-k}I_{2}^{k}F_{k}^{B}(u) ,\\&&F_{k}^{B}(u)=(-1)^{k}
 \binom{\ell}{k}\frac{1}{u^{\ell+k+2}} ,\label{bare}
\end{eqnarray}
we realize that the main task is the extraction of the singularities
from the  \emph{bare} distributions $F_{k}^{B}\sim u^{-n}, n \in
\textrm{N}$. This can be done by shifting the integer $n$ by some
infinitesimal amount $\epsilon$ i.e. $u^{\epsilon-n}$. Considering
some smooth test function $f(u), u\in \textrm{R}^{+}$ and using a
Laplace transformation ($ f(u)=\int^{\infty}_{0}ds
e^{-us}\hat{f}(s)$) we can write
\begin{eqnarray}
  && \int^{\infty}_{0}\frac{f(u)}{u^{n-\epsilon}}du=
  \int^{\infty}_{0}ds
  \hat{f}(s)s^{n-\epsilon-1}\Gamma(\epsilon-n+1)\\
  && =\frac{\partial^{n-1}}{\partial u^{n-1}}f(u)|_{u=0}
  \left(\frac{1}{\epsilon(n-1)!}+\textnormal{reg.
  part}\right) .
\end{eqnarray}
So we see that the singular part of our distribution is the $n-1$
order derivative of the delta function
\begin{equation}\label{sd}
    \left[\frac{1}{u^{n-\epsilon}}\right]_{sing}=
    \frac{1}{\epsilon}\frac{(-1)^{n-1}}{(n-1)!}\delta^{(n-1)}(u) .
\end{equation}
The next step is to connect this $\epsilon$ shifting with some
\emph{gauge invariant} scheme  such as dimensional regularization
for our bare correlator formed by the set of distributions
(\ref{bare}) and the Ward identities (\ref{Ward}). Finally we will
preserve the Ward identity of current conservation necessary for
gauge invariance and extract the violation of tracelessness for this
case.

 For doing this note that the set
$F^{B}_{k}$ satisfies  the conservation and tracelessness conditions
for $d=3\,\,(D=4)$ if we understand them as analytic functions of
$u$. Using (\ref{dv}) and (\ref{tr}) for $d=3$ we can see that
\begin{eqnarray}
  && (Div_{\ell}F^{B})_{k}^{d=3}=0,\quad(Tr_{\ell}F^{B})^{d=3}_{k}=0
  .
\end{eqnarray}
On the other hand it is easy to see that we can satisfy Ward
identities for general $d$ if we regularize the bare distributions
in the following way
\begin{equation}\label{regu}
    F^{R}_{k}=(-1)^{k}
 \binom{\ell}{k}\frac{1}{u^{\ell+k+d-1}},
\end{equation}
which depends analytically on $d$ with a pole at $d=3$. Actually we
need to check only the conservation condition (\ref{dv}) because
tracelessness (\ref{tr}) does not include derivatives and dimension.
Indeed
\begin{eqnarray}
&& (Div_{\ell}F^{R})_{k}=(\ell-k)(u+1)
  \partial_{u}F^{R}_{k}+(k+1)u(u+2)\partial_{u}F^{R}_{k+1}\nonumber\\
  &&+(\ell-k)(\ell+d+k)F^{R}_{k}+(k+1)(\ell+d+k+1)(u+1)F^{R}_{k+1}=0 ,\label{dvr}\\
  &&(Tr_{\ell}F^{R})_{k}=(\ell-k)(\ell-k-1)F^{R}_{k}
  +2(k+1)(\ell-k-1)(u+1)
F^{R}_{k+1}\nonumber\\&&\quad\quad\quad\quad\quad+(k+2)(k+1)u(u+2)F^{R}_{k+2}=0
,\label{tr1}
\end{eqnarray}
 hold analytically in $d$. So we can just put in (\ref{dvr}),(\ref{tr1}) $d=3-\epsilon$
 and say that we constructed the regularized Ward identities.

 Then the procedure is more or less standard. We can split (\ref{regu})
for $d=3-\epsilon$
 in a singular and renormalized parts using (\ref{sd})
 \begin{equation}\label{ren}
    F^{R}_{k}(u)=
 \binom{\ell}{k}\left(\frac{1}{\epsilon}\frac{(-1)^{\ell+1}}
 {(\ell+k+1)!}\delta^{(\ell+k+1)}(u)+f_{k}(u)\right) +F^{Ren}_{k}(u)
 ,
 \end{equation}
where we included also  a set of finite distributions $f_{k}(u)$
(without $\epsilon$ pole)  to describe the finite renormalization
freedom. Analyzing (\ref{ren}) we can say that our singular part
corresponds to the local counterterms of the effective action
because each is proportional to a derivative of the delta function.
On the other side the renormalized correlation function formed by
$F^{Ren}_{k}(u)$ will  on the quantum level get the same trace as a
 subtracted singular part but with opposite sign because the
regularized expression is traceless and conserved. So we can insert
the subtraction parts
\begin{equation}\label{singw}
    F^{S}_{k}(u)=
 \binom{\ell}{k}\left(\frac{1}{\epsilon}\frac{(-1)^{\ell+1}}
 {(\ell+k+1)!}\delta^{(\ell+k+1)}(u)+f_{k}(u)\right)
\end{equation}
in the regularized current conservation Ward identity (\ref{dvr})
for $d=3-\epsilon$ and obtain equations for the set of $f_{k}(u)$
after sending $\epsilon\rightarrow 0$ . Using the relation
$u\delta^{(n)}(u)=-n\delta^{(n-1)}(u)$ we obtain
\begin{eqnarray}
  &&(u+1)f'_{k}+u(u+2)f'_{k+1}+(\ell+3+k)f_{k}+(\ell+4+k)(u+1)f_{k+1}\nonumber\\
  &&=(-1)^{\ell+1}\frac{\delta^{(\ell+k+2)}}{(\ell+k+2)!} , \quad\quad
  k=0,1,\dots\ell-1  .\label{deltaf}
\end{eqnarray}

For finding a solution we introduce a suitable ansatz
\begin{eqnarray}\label{ans}
    f_{k}(u)=\sum^{k+1}_{p=0}g^{p}_{k}\frac{\delta^{(\ell+k+1-p)}(u)}{(\ell+k+1-p)!}
    .
\end{eqnarray}
Substituting (\ref{ans}) in (\ref{deltaf}) we obtain the following
recursion equations for the unknown coefficients $g^{p}_{k}$
\begin{eqnarray}
  && g^{0}_{k+1}-g^{0}_{k}=-\frac{1}{\ell+k+2} ,\label{inc}\\
  &&
  \frac{g^{p+1}_{k+1}}{(\ell+k+1-2p)_{p+1}}-\frac{g^{p+1}_{k}}{(\ell+k-2p)_{p+1}}\nonumber\\
  &&=-\frac{p+1}{(\ell+k-2p)_{p+2}}\left(g^{p}_{k+1}-g^{p}_{k}\right),
   \quad\quad p=1,2,\dots k  ,\label{rec}\\
  && g^{k+1}_{k+1}-g^{k+1}_{k}=0 .\label{cutc}
\end{eqnarray}
Now note that the following nontrivial expression solves the
recursion (\ref{rec}) with initial condition (\ref{inc}) (see the
proof in Appendix C)
\begin{eqnarray}
  &&
  g^{p}_{k}=\frac{(-1)^{p}(p-1)!}{2^{p}(\ell+k+2-p)_{p}}\nonumber\\
  &&+\sum^{[p/2]}_{n=0}\frac{n!}{2^{n}}\binom{p}{n}
  \binom{p-n}{n}(\ell+k-2(p-n-1))_{p-2n}\Delta_{p-n} ,\label{sol}
\end{eqnarray}
where $\Delta_{1},\dots\Delta_{\ell}$ is a set of $\ell$ unknown
constants. Substituting this solution in the so-called cutting
conditions (\ref{cutc}) we obtain $\ell$ linear equations for the
unknown constants $\Delta_{n}$ in triangular form
\begin{eqnarray}
  &&
  \frac{(-1)^{k}}{(\ell+k+2)!}+\sum^{[k/2]}_{n=0}
  \frac{2^{k+1-n}\Delta_{k+1-n}}{n!}\binom{\ell}{k-2n}=0 .
\end{eqnarray}
So we prove that there is a consistent solution for the equations
(\ref{deltaf}) and we manage that the singular part of the
correlation function with an appropriate choice of the finite part
$f_{k}$ does not violate the gauge Ward identity. Of course this
solution  violates tracelessness of the loop function due to the
existence of the finite part and we can say that we observed a
higher spin version of the trace anomaly. The important point is the
following: Even after the violation of tracelessness our theory is
still in the framework of Fronsdal's \cite{Frons} gauge invariance
for massless but only double traceless gauge fields. In this
formulation the conservation condition for the currents with the
nonzero trace looks like
\begin{equation}\label{fcc}
\nabla^{\mu}\frac{\partial}{\partial
a^{\mu}}J^{(\ell)}(a;z)=O(a^{2}) .
\end{equation}
This corresponds to the usual conservation for the part of the
currents expanded in $I_{1},I_{2}$ bitensors  which we actually
checked here.
 Finally note that because the initial conditions
(\ref{inc}) include only the difference between pairs of neighboring
variables $g^{0}_{k}$ we have as one  degree of freedom an arbitrary
$g^{0}_{0}$. This we can interpret as an arbitrary renormalization
point.

\section{Renormalization and RG equations}
Now we return to (\ref{ren})
 \begin{equation}\label{ren1}
    F^{R}_{k}(u)=
 \binom{\ell}{k}\left(\frac{1}{\epsilon}\frac{(-1)^{\ell+1}}
 {(\ell+k+1)!}\delta^{(\ell+k+1)}(u)+f_{k}(u)\right) +F^{Ren}_{k}(u)
 ,
 \end{equation}
where the finite renormalization part is (\ref{ans})
\begin{eqnarray}\label{ans1}
    f_{k}(u)=\sum^{k+1}_{p=0}g^{p}_{k}\frac{\delta^{(\ell+k+1-p)}(u)}{(\ell+k+1-p)!}
    .
\end{eqnarray}
The set of unknown constants $g^{p}_{k}$ we can find from the system
of equations (\ref{inc})-(\ref{cutc}).

We note that because the initial conditions (\ref{inc}) include only
the difference between pairs of neighboring variables $g^{0}_{k}$ we
have as one  degree of freedom an arbitrary
$g^{0}_{0}=(-1)^{\ell+1}\mu$. The parameter $\mu$ we can interpret
as an arbitrary renormalization point. The important point here is
the following: \emph{Equation (\ref{inc}) leads to the special
dependence of  all the $g^{0}_{k}$ from $\mu$ }
\begin{eqnarray}\label{aaa}
&&g^{0}_{k}=(-1)^{\ell+1}\mu+ \tilde{g}^{0}_{k} ,\quad \tilde{g}^{0}_{0}=0\\
&&f_{k}(u)=(-1)^{\ell+1}\mu\frac{\delta^{(\ell+k+1)}(u)}{(\ell+k+1)!}
+\sum^{k+1}_{p=1}\tilde{g}^{p}_{k}\frac{\delta^{(\ell+k+1-p)}(u)}{(\ell+k+1-p)!}
\end{eqnarray}
So we obtain the following dependence of the renormalized and
singular parts on $\mu$
\begin{equation}\label{ren2}
    F^{R}_{k}(u)=
 \binom{\ell}{k}\left(\left[\frac{1}{\epsilon}+\mu\right]\frac{(-1)^{\ell+1}}
 {(\ell+k+1)!}\delta^{(\ell+k+1)}(u)+\tilde{f}_{k}(u;\tilde{g}^{p}_{k})\right) +F^{Ren}_{k}(u,\mu)
 ,
 \end{equation}
Then in a standard way we can derive the RG equations from the $\mu$
independence of the regularized set $\frac{d}{d\mu}F^{R}_{k}(u)=0$
\begin{equation}\label{bbb}
    \frac{d}{d\mu}F^{Ren}_{k}(u,\mu)=\binom{\ell}{k}\frac{(-1)^{\ell}}
 {(\ell+k+1)!}\delta^{(\ell+k+1)}(u)
\end{equation}
Using this and our integration procedure from the next section we
can express the RG equations by the effective action
\section{Singular parts of the effective action and integration}
The singular part of the two point function of the higher spin
currents in $AdS_{d+1}$ space can be expressed in the following form
\begin{eqnarray}
  && K(a,c;z_{1},z_{2})=\sum^{\ell}_{k=0}I_{1}^{\ell-k}(a,c)I^{k}_{2}(a,c)F^{S}_{k}(u;\mu) \\
  && F^{S}_{k}(u;\mu)=
 \binom{\ell}{k}\left(\left[\frac{1}{\epsilon}+\mu\right]\frac{(-1)^{\ell+1}}
 {(\ell+k+1)!}\delta^{(\ell+k+1)}(u)+\tilde{f}_{k}(u;\tilde{g}^{p}_{k})\right)\label{count1}\\
 &&\tilde{f}_{k}(u;\tilde{g}^{p}_{k})=\sum^{k+1}_{p=1}\tilde{g}^{p}_{k}\frac{\delta^{(\ell+k+1-p)}(u)}{(\ell+k+1-p)!}
\end{eqnarray}
We are going to investigate the following integral
\begin{eqnarray}
   &&
   h^{(\ell)}(a;z_{1}) \ast_{a_{1}}K(a,c;z_{1},z_{2})\ast_{c_{2}} h^{(\ell)}(c;z_{2}) \\
  &&\ast_{a_{1}}=\int\sqrt{g}d^{4}z_{1}\left[\overleftarrow{\partial}^{\mu}_{a}\overrightarrow
  {\partial}^{a}_{\mu}\right]^{\ell}=\int\sqrt{g}d^{4}z_{1}\hat{\ast}_{a}
\end{eqnarray}

First of all we have to express $\delta^{(n)}(u)$ as a combination
of the covariant derivatives of the four dimensional delta function
in the general coordinate system. It is possible because the
parameter $u$ is an $AdS$ invariant variable. We start from the
covariant delta function in the curved space with the metric
$g_{\mu\nu}(z)$ and invariant measure $d\mu(z)=\sqrt{g}d^{4}z$
\begin{equation}\label{delta}
    \frac{\delta_{(4)}(z-a)}{\sqrt{g(z)}} ,\quad\quad\quad\quad \int\delta_{(4)}(z-a)d^{4}z=1
\end{equation}
In the polar coordinate system (see Appendix A) the invariant
measure is
\begin{equation}\label{m}
d\mu(z)=\sqrt{g}d^{4}z=u(u+2)dud\Omega_{3}
\end{equation}
Therefore we can define
\begin{eqnarray}
  && \frac{\delta_{(4)}(z-z_{pole})}{\sqrt{g(z)}}=\frac{\delta(u)}{u(u+2)\Omega_{3}}
  =-\frac{\delta^{(1)}(u)}{(u+2)\Omega_{3}} \\
  && u\delta^{(1)}(u)=-\delta(u)
\end{eqnarray}
Applying (\ref{lm3}) we can derive
\begin{equation}\label{rec1}
    -(u+2)\Box\frac{\delta^{(n)}(u)}{u+2}
    =2n\delta^{(n+1)}(u)+[2-n(n+1)]\delta^{(n)}(u)
\end{equation}
which can be formulated as a recursion for the object
$\phi_{n}(u)=\frac{\delta^{(n)}}{u+2}$
\begin{eqnarray}
  && \phi_{n+1}(u)=-\hat{D}_{n}\phi_{n}(u) \label{rec2}\\
  && \hat{D}_{n}=\frac{1}{2n}\left[\Box+2-n(n+1)\right]
\end{eqnarray}
So it is easy to see that because
$\phi_{1}(u)=-\frac{\delta_{(4)}(z-z_{pole})}{\sqrt{g(z)}}\Omega_{3}$
we can express the solution of (\ref{rec2}) in the form
\begin{equation}\label{sol1}
\phi_{n+1}(u)=(-1)^{n+1}\Omega_{3}\left\{\prod^{n}_{m=1}\hat{D}_{m}\right\}\frac{\delta_{(4)}(z-z_{pole})}{\sqrt{g(z)}}
\end{equation}
Then using $\delta^{(n)}(u)=2\phi_{n}-n\phi_{n-1}$ we obtain the
final conversion formula
\begin{equation}\label{sol2}
\delta^{(n)}(u)=(-1)^{n}\Omega_{3}\left\{2\hat{D}_{n-1}
+n\right\}\left\{\prod^{n-2}_{m=1}\hat{D}_{m}\right\}\frac{\delta_{(4)}(z-z_{pole})}{\sqrt{g(z)}}
\end{equation}

Now we concentrate on the singular and $\mu$ dependent part of
(\ref{count1})
\begin{equation}\label{count2}
    [\frac{1}{\epsilon}+\mu]h^{(\ell)}(a;z_{1}) \ast_{a_{1}} \sum^{\ell}_{k=0}I_{1}^{\ell-k}(a,c)I^{k}_{2}(a,c)(-1)^{\ell+1}
 \binom{\ell}{k}\frac{\delta^{(\ell+k+1)}(u)}
 {(\ell+k+1)!}\ast_{c_{2}} h^{(\ell)}(c;z_{2})\quad\quad
\end{equation}
Admitting that our higher spin gauge field is transversal and
traceless and integrating partially we obtain
\begin{eqnarray}
  &&[\frac{1}{\epsilon}+\mu] (-1)^{\ell+1}Z^{\ell}h^{(\ell)}(a;z_{1}) \ast_{a_{1}}I_{1}^{\ell}\delta^{(\ell+1)}(u)
 \ast_{c_{2}} h^{(\ell)}(c;z_{2}),  \label{act}\\
  &&Z^{\ell}=\frac{1}{(\ell+1)}\sum^{\ell}_{k=0}
  \frac{(-1)^{k}}{(\ell-k)!(\ell+2)_{k}}=\frac{1}{(2\ell+1)\ell!}
\end{eqnarray}
Next we describe the way to take one integral in (\ref{act}).
Considering the following expression
\begin{eqnarray}
  && \tilde{K}(a;z_{1})=(-1)^{\ell+1}Z^{\ell}I_{1}^{\ell}\delta^{(\ell+1)}(u)
 \ast_{c_{2}} h^{(\ell)}(c;z_{2}) ,
\end{eqnarray}
we can using a conformal transformation fix the point $z_{1}$ as a
pole for the coordinate system $z_{2}$. In this case we can insert
directly conversion formula (\ref{sol2}) and obtain
\begin{equation}
  \tilde{K}(a;z_{pole})= Z^{\ell}\Omega_{3}I_{1}^{\ell}\left\{2\hat{D}_{\ell}
+(\ell+1)\right\}\left\{\prod^{\ell-1}_{m=1}\hat{D}_{m}\right\}\frac{\delta_{(4)}(z_{2}-z_{pole})}
{\sqrt{g(z_{2})}}\ast_{c_{2}} h^{(\ell)}(c;z_{2})\quad\quad
\end{equation}
Remembering the following formula (for transverse and traceless
$h^{(\ell)}$)
\begin{equation}\label{rel1}
    \Box I_{1}^{\ell}\hat{\ast}_{c_{2}}h^{(\ell)}(c;z_{2})=I_{1}^{\ell}
    \hat{\ast}_{c_{2}}\left\{\Box+\ell\right\}h^{(\ell)}(c;z_{2})
\end{equation}
we obtain finally
\begin{equation}
\tilde{K}(a;z_{pole})=
Z^{\ell}\Omega_{3}\left\{(a^{\mu}c_{\mu})^{\ell}\left\{2\hat{D}_{\ell}
+(\ell+2)\right\}\prod^{\ell-1}_{m=1}\left[\hat{D}_{m}+\frac{\ell}{2m}\right]\hat{\ast}_{c}
h^{(\ell)}(c;z)\right\}_{z=z_{pole}}
\end{equation}
With the help of this formula and (\ref{ren2}) the singular part of
the one loop effective action (\ref{count2}) for transversal and
traceless external spin $\ell$ field can be expressed in the
following local form
\begin{eqnarray}
  && W^{\ell}_{Sing}(h^{(\ell)},\mu)=\left[\frac{1}{\epsilon}+\mu\right]Z^{\ell}\Omega_{3}\int
   \sqrt{g}d^{4}z h^{(\ell)}_{\mu_{1}\dots\mu_{\ell}}K^{\ell}
   (\Box)h^{(\ell)\mu_{1}\dots\mu_{\ell}} \label{efa1}\\
  && K^{\ell}(\Box)=\left\{2\hat{D}_{\ell}
+(\ell+2)\right\}\prod^{\ell-1}_{m=1}\left[\hat{D}_{m}+\frac{\ell}{2m}\right]\label{kop}
\end{eqnarray}
Then from (\ref{bbb}) the  scale anomaly (integrated trace anomaly)
for the renormalized effective action comes out as
\begin{eqnarray}
  && \frac{d}{d\mu}W^{\ell}_{Ren}(h^{(\ell)},\mu)=
  -\frac{d}{d\mu}W^{\ell}_{Sing}(h^{(\ell)},\mu)\nonumber\\
  &&=-Z^{\ell}\Omega_{3}\int
   \sqrt{g}d^{4}z h^{(\ell)}_{\mu_{1}\dots\mu_{\ell}}K^{\ell}
   (\Box)h^{(\ell)\mu_{1}\dots\mu_{\ell}}
\end{eqnarray}

In the case of $\ell=2$ this integral should be proportional to the
integrated square of the gravitational Weyl tensor linearized in the
$AdS_{4}$ background (see \cite{BD} and ref. there)
\begin{eqnarray}
  && C^{\mu\nu}_{\lambda\rho}(G)C^{\lambda\rho}_{\mu\mu}(G)=
  R^{\mu\nu}_{\lambda\rho}(G)R^{\lambda\rho}_{\mu\nu}(G)
  -2R^{\mu\nu}(G)R_{\mu\nu}(G)+\frac{1}{3}R(G)R(G)\label{c22}\\
  &&G_{\mu\nu}=g_{\mu\nu}+h^{(2)}_{\mu\nu}\quad ,\quad
  \nabla^{\mu}h^{(2)}_{\mu\nu}=h^{(2)\mu}_{\mu}=0
\end{eqnarray}
For traceless and transversal $h^{(2)}_{\mu\nu}$ in an $AdS_{4}$
background (we put as before $L$=1) we have
\begin{eqnarray}
  && R^{\mu\nu}_{\lambda\rho}(G)=R^{\mu\nu}_{\lambda\rho}(h^{(2)})=
  2\nabla^{[\mu}\nabla_{[\lambda}h^{(2)\nu]}_{\rho]}-2\delta^{[\mu}_{[\lambda}h^{(2)\nu]}_{\rho]} \\
  && R^{\mu}_{\lambda}(h^{(2)})=\frac{1}{2}\Box
  h^{(2)\mu}_{\lambda}+ h^{(2)\mu}_{\lambda}\quad ,\quad R(h^{(2)})=0
\end{eqnarray}
and straightforward calculations lead to
\begin{equation}\label{c21}
    \int\sqrt{g}d^{4}z C^{\mu\nu}_{\lambda\rho}(h^{(2)})C^{\lambda\rho}_{\mu\mu}(h^{(2)})
    =\frac{1}{2}\int\sqrt{g}d^{4}z
    h^{(2)}_{\mu\nu}\left[\Box^{2}+6\Box+8\right]h^{(2)\mu\nu} .
\end{equation}
Then we can evaluate (\ref{kop}) for $\ell=2$ and obtain
\begin{equation}\label{kop2}
    K^{2}(\Box)=\frac{1}{4}\left[\Box^{2}+6\Box+8\right] .
\end{equation}
So we see that
\begin{equation}\label{finfor}
     W^{2}_{Sing}(h^{(\ell)},\mu)=\left[\frac{1}{\epsilon}+\mu\right]
     \frac{Z^{2}\Omega_{3}}{2}\int\sqrt{g}d^{4}z
C^{\mu\nu}_{\lambda\rho}(h^{(2)})C^{\lambda\rho}_{\mu\nu}(h^{(2)})
\end{equation}

 Finally note that if our external higher spin field is
\emph{on-shell} we can replace all Laplacians using the equation of
motion
\begin{eqnarray}
&&[\Box +\ell ]h^{(\ell )}=\Delta _{\ell }(\Delta _{\ell
}-3)h^{(\ell )},
\label{ph1} \\
&&\Box _{a}h^{(\ell )}=\nabla ^{\mu }\frac{\partial } {\partial
a^{\mu }}
h^{(\ell )}=0,  \label{pd2} \\
&&\Delta _{\ell }=\ell +1.  \label{ph3}
\end{eqnarray}

\section*{Conclusions}
In this article we considered the two point correlation function for
traceless conserved higher spin currents in $AdS_{4}$. Using a kind
of dimensional regularization scheme we defined the correct
renormalization procedure for the one loop diagram corresponding to
this correlator and investigated the Ward identities. We have shown
that extracting the delta function singularities we can define the
renormalization in a gauge invariant way and obtain the violation of
tracelessness. This means that we observed a \emph{trace anomaly}
for higher spin currents. This result was used for the derivation of
the one loop anomalous effective action of the conformal scalar in
$AdS$ space that is minimally coupled to the higher spin external
field or for the investigation of the one-loop renormalized
propagators for the higher spin conformal gauge fields with
linearized interaction with the conformal scalar.

\subsection*{Acknowledgements}
\quad This work is supported in part by the German
Volkswagenstiftung. The work of R.~M. was supported by DFG (Deutsche
Forschungsgemeinschaft) and in part by the INTAS grant \#03-51-6346.

\section*{Appendix A}
\setcounter{equation}{0}
\renewcommand{\theequation}{A.\arabic{equation}}
The Euclidian $AdS_{d+1}$ metric
\begin{equation}\label{eads}
    ds^{2}=g_{\mu \nu }(z)dz^{\mu }dz^{\nu
}=\frac{1}{(z^{0})^{2}}\delta _{\mu \nu }dz^{\mu }dz^{\nu }
\end{equation}
can be realized as an induced metric for the hypersphere defined by
the following embedding procedure in $d+2$ dimensional Minkowski
space
\begin{eqnarray}
  && X^{A}X^{B}\eta_{AB}=-X_{-1}^{2}+X_{0}^{2}+\sum^{d}_{i=1}X_{i}^{2}=-1 ,\\
  && X_{-1}(z)=\frac{1}{2}\left(\frac{1}{z_{0}}+\frac{z_{0}^{2}
  +\sum^{d}_{i=1}z^{2}_{i}}{z_{0}}\right) ,\\
  && X_{0}(z)=\frac{1}{2}\left(\frac{1}{z_{0}}-\frac{z_{0}^{2}
  +\sum^{d}_{i=1}z^{2}_{i}}{z_{0}}\right) ,\\
  && X_{i}(z)=\frac{z_{i}}{z_{0}} .
\end{eqnarray}
Using this embedding rules we can realize that the geodesic distance
$\zeta(z,w)$ is just an $SO(1,d+1)$ invariant scalar product
\begin{equation}\label{gd}
    -X^{A}(z)Y^{B}(w)\eta_{AB}=\frac{1}{2z_{0}w_{0}}\left(2z_{0}w_{0}
    +\sum^{d}_{\mu=0}(z-w)^{2}_{\mu}\right)=\zeta ,
\end{equation}
and therefore can be realized by a hyperbolic angle. Indeed we can
introduce another embedding
\begin{eqnarray}\label{hyp}
    &&X_{-1}(\Theta,\omega_{\mu})=\cosh{\Theta},\\
   && X_{\mu}(\Theta,\omega_{\mu})=\sinh{\Theta}\,\omega_{\mu}\quad,\quad\quad
    \sum^{d}_{\mu=0}\omega_{\mu}=1 ,\\
    &&ds^{2}=d\Theta^{2}+\sinh^{2}{\Theta}\, d\Omega_{d} .
\end{eqnarray}
In these coordinates the geodesic distance between an arbitrary
point $X^{A}(\Theta,\Omega_{\mu})$ and the pole of the hypersphere
$Y^{A}(\Theta=0,\omega_{\mu})$ is simply
\begin{equation}\label{hd}
    \zeta= -X^{A}Y^{B}\eta_{AB}=\cosh{\Theta} .
\end{equation}
Therefore the invariant measure is expressed as
\begin{equation}\label{invv}
    \sqrt{g}d\Theta d\Omega_{d}=(\sinh\Theta)^{d}d\Theta
    d\Omega_{d}=(\zeta^{2}-1)^{\frac{d-1}{2}}d\zeta d\Omega_{d} .
\end{equation}
So we see that the integration measure for $d=3$ ($D=d+1=4$) will
cancel one order of $(\zeta-1)^{-n}$ in short distance singularities
and we have to count  the singularities starting from
$(\zeta-1)^{-2}$.

 In this article we use the following rules and relations for
$\zeta(z,z')$, $I_{1a}$, $I_{2c}$ and the bitensorial basis
$\{I_{i}\}^{4}_{i=1}$
\begin{eqnarray}
  && \Box\zeta=(d+1)\zeta ,\quad \nabla_{\mu}\partial_{\nu}\zeta=g_{\mu\nu}\zeta ,
  \quad g^{\mu\nu}\partial_{\mu}\zeta\partial_{\nu}\zeta=\zeta^{2}-1 ,\\
  &&   \partial_{\mu}\partial_{\nu'}\zeta
  \nabla^{\mu}\zeta=\zeta\partial_{\nu'}\zeta ,\quad
  \partial_{\mu}\partial_{\nu'}\zeta \nabla^{\mu}\partial_{\mu'}\zeta
  =g_{\mu'`\nu'}+\partial_{\mu'}\zeta\partial_{\nu'}\zeta ,\\
&&\nabla_{\mu}\partial_{\nu}\partial_{\nu'}\zeta \nabla^{\mu}\zeta
  =\partial_{\nu}\zeta\partial_{\nu'}\zeta ,\quad
  \nabla_{\mu}\partial_{\nu}\partial_{\nu'}\zeta
  =g_{\mu\nu}\partial_{\nu'}\zeta ,\\
&&\frac{\partial}{\partial a^{\mu}}I_{1a}\frac{\partial}{\partial
a_{\mu}}I_{1a}=\zeta^{2}-1 ,\quad \frac{\partial}{\partial
a^{\mu}}I_{1}\frac{\partial}{\partial
a_{\mu}}I_{1a}=\zeta I_{2c} ,\\
&&\frac{\partial}{\partial a^{\mu}}I_{1}\frac{\partial}{\partial
a_{\mu}}I_{1}=c^{2}_{2}+ I_{2c}^{2} , \, \frac{\partial}{\partial
a^{\mu}}I_{1}\frac{\partial}{\partial a_{\mu}}I_{2}=\zeta I_{2c}^{2}
,\,\Box_{a}I_{4}=2(d+1)c^{2}_{2} ,\\
 &&\frac{\partial}{\partial
a^{\mu}}I_{2}\frac{\partial}{\partial
a_{\mu}}I_{2}=(\zeta^{2}-1)I_{2c}^{2} ,\quad
\Box_{a}I_{3}=2(d+1)I_{2c}^{2}+2c^{2}_{2}(\zeta^{2}-1) ,\\
&&\nabla^{\mu}\frac{\partial}{\partial a^{\mu}}I_{1}=(d+1)I_{2c}
,\,\nabla^{\mu}\frac{\partial}{\partial a^{\mu}}I_{2}=(d+2)\zeta
I_{2c},\quad\nabla^{\mu} I_{1}\partial_{\mu}
\zeta=I_{2} ,\\
&&\nabla^{\mu}\frac{\partial}{\partial
a^{\mu}}I_{3}=4I_{1}I_{2c}+2(d+2)\zeta c^{2}_{2}I_{1a}
,\quad\nabla^{\mu} I_{2}\partial_{\mu} \zeta=2\zeta I_{2}
,\end{eqnarray}
\begin{eqnarray}
&&\frac{\partial}{\partial a_{\mu}} I_{1}\partial_{\mu} \zeta=\zeta
I_{2c} ,\quad \frac{\partial}{\partial a_{\mu}} I_{2}\partial_{\mu}
\zeta=(\zeta^2-1) I_{2c} ,\,\frac{\partial}{\partial a_{\mu}}
I_{1}\nabla_{\mu} I_{1}=I_{1} I_{2c} ,\,\,\,\,\,\\
&&\frac{\partial}{\partial a_{\mu}} I_{1}\nabla_{\mu}
I_{2}=I_{2c}\left(\zeta I_{1}+I_{2}\right)+c^{2}_{2}I_{1a}
,\frac{\partial}{\partial a_{\mu}} I_{2}\nabla_{\mu}
I_{1}=I_{2c}I_{2} ,\\
&&\frac{\partial}{\partial a_{\mu}} I_{2}\nabla_{\mu} I_{2}=2\zeta
I_{2c}I_{2} ,\quad \nabla^{\mu} I_{1}\nabla_{\mu}
I_{1}=a^{2}_{1}I_{2c} ,\quad \Box I_{1}=I_{1} ,\\
&&\nabla^{\mu} I_{1}\nabla_{\mu} I_{2}=I_{2}I_{1}+ a^{2}_{1}\zeta
I_{2c} ,\quad \Box I_{2}=(d+2)I_{2}+2\zeta
I_{1} ,\quad\\
&&\nabla^{\mu} I_{2}\nabla_{\mu} I_{2}=I_{2}^{2}+2\zeta
I_{1}I_{2}+a^{2}_{1}I_{2c}^{2}\zeta^{2}+c^{2}_{2}I_{1a}^{2}
,\quad\nabla^{\mu} I_{2}\partial_{\mu}\zeta=2\zeta I_{2} ,\\
&&a^{\mu}\nabla_{\mu}I_{1a}=a^{2}\zeta ,\quad
a^{\mu}\nabla_{\mu}I_{2c}=I_{1},\quad
a^{\mu}\nabla_{\mu}I_{1}=a^{2}I_{2c}
,\\&&a^{\mu}\nabla_{\mu}I_{2}=a^{2}\zeta I_{2c}+I_{1a}I_{1}
,\quad\nabla^{\mu} I_{1}\partial_{\mu}\zeta=I_{2}.
\end{eqnarray}

Using these relations we can derive ($
F'_{k}:=\frac{\partial}{\partial
  \zeta}F_{k}(\zeta)$)
\begin{itemize}
  \item Divergence map
\begin{eqnarray}
  && \nabla^{\mu}_{1}\frac{\partial}{\partial a^{\mu}}\Psi^{\ell}[F]=
  I_{2c}\Psi^{\ell-1}[Div_{\ell}F]+O(c^{2}_{2}) , \label{div}\\
  && (Div_{\ell}F)_{k}=(\ell-k)\zeta
  F'_{k}+(k+1)(\zeta^{2}-1)F'_{k+1}\nonumber\\
  &&+(\ell-k)(\ell+d+k)F_{k}+(k+1)(\ell+d+k+1)\zeta F_{k+1} .\label{dv}
  \end{eqnarray}
  \item Trace map
\begin{eqnarray}
    &&\Box_{a}\Psi^{\ell}[F]=I^{2}_{2c}\Psi^{\ell-2}
    [{Tr_{\ell}F}]+O(c^{2}_{2}) ,\label{tracemap}\\
&&(Tr_{\ell}F)_{k}=(\ell-k)(\ell-k-1)F_{k} +2(k+1)(\ell-k-1)\zeta
F_{k+1}\nonumber\\&&\quad\quad\quad\quad\quad+(k+2)(k+1)(\zeta^{2
}-1)F_{k+2} .\label{tr}
\end{eqnarray}
  \item Laplacian map
  \begin{eqnarray}
  && \Box_{1} \Psi^{\ell}[F]=\Psi^{\ell}[Lap_{\ell}F]+O(a^{2}_{1},c^{2}_{2}) , \label{lm1}\\
  &&(Lap_{\ell}F)_{k}=(\zeta^{2}-1)F''_{k}+(d+1+4k)\zeta
  F'_{k}+[\ell+k(d+2\ell-k)]F_{k}\nonumber\\&&+2\zeta(k+1)^{2}
  F_{k+1}+2(\ell-k+1)F'_{k-1},\label{lm2}\\
  &&\Box F_{k}(\zeta)=(\zeta^{2}-1)F''_{k}+(d+1)\zeta F'_{k}.\quad\label{lm3}\end{eqnarray}
  \item Gradient map
  \begin{eqnarray}
&&(a\cdot\nabla)_{1}\Psi^{\ell}[F]=I_{1a}\Psi^{\ell}[Grad_{\ell}F]+ O(a^{2}_{1}) ,\label{grad1}\\
  && (Grad_{\ell}F)_{k}=F'_{k}+(k+1)F_{k+1} .\label{grad2}
\end{eqnarray}
\item Bigradient map
\begin{eqnarray}
&&\Psi^{(\ell+1)}[G]=(a\cdot\nabla_{1})(c\cdot\nabla_{2})
\Psi^{(\ell)}[F]+ O(a^{2}_{1}, c^{2}_{2}) \label{bimap1} ,\\
&&G_{k}=F''_{k-1}(\zeta)+(2k+1)F'_{k}(\zeta)+(k+1)^{2}F_{k+1}\label{bimap2}
\end{eqnarray}
\end{itemize}
\section*{Appendix B}
\setcounter{equation}{0}
\renewcommand{\theequation}{B.\arabic{equation}}

Considering (\ref{mix1}) as an ansatz we must solve (see
(\ref{grad1})) the system of differential equations
\begin{equation}\label{b1}
   F_{k,sing}^{(\ell)}\hat{=}\,G''_{k-1}(\zeta)+(2k+1)G'_{k}(\zeta)+(k+1)^{2}G_{k+1}
\end{equation}
($\hat{=}$ means: modulo regular terms). Since $G_{k}=0$ for $k\geq
\ell$, this system is solved recursively, starting with $k=\ell$,
and lowering $k$ step by step. The arbitrary polynomials of $\zeta$
obtained by integration are discarded since they are regular. The
solution $\left\{G_{k}\right\}^{\ell-1}_{k=0}$ is therefore unique.

From (\ref{ms1}) we obtain as solution
\begin{equation}\label{b2}
    G_{k}\hat{=}\,\sum^{\ell+1}_{m=2}\sum^{\ell-k}_{n=1}
    \frac{a^{(\ell)}_{k+n,m}}{(m+k-1)_{n+1}}P_{n-1}(k)(\zeta-1)^{-m-k+1}\,,
\end{equation}
where $P_{n}(x)$ are polynomials of degree $n$ defined by
\begin{equation}\label{b3}
P_{n}(x) =
\frac{d}{dx}(x+1)_{n+1}=(x+1)_{n+1}[\psi(x+n+2)-\psi(x+1)].
\end{equation}
By integration of a singular term\, $(\zeta-1)^{-2}$ we obtain a
regular term $(\zeta-1)^{-1}$. This has happened in (\ref{b2}): The
term $m=2$ is regular. Discarding this "leading regular term" and
renumbering the sum we get
\begin{equation}\label{b4}
    G_{k}\hat{=}\,\sum^{\ell}_{m=2}\sum^{\ell-k}_{n=1}
    \frac{a^{(\ell)}_{k+n,m+1}}{(m+k)_{n+1}}P_{n-1}(k)(\zeta-1)^{-m-k}\,,
\end{equation}

For $k=0$ we obtain a differential equation from (\ref{b1}) which
acts as an integrability constraint
\begin{equation}\label{b5}
    F^{(\ell)}_{0,sing}-G'_{0}-G_{1} \hat{=}\,0 .
\end{equation}
Inserting the expansion (\ref{mix1}) and (\ref{b2}) in (\ref{b4}) we
get from (\ref{b5})
\begin{equation}\label{b6}
    \sum^{\ell}_{k=0}\frac{k!}{(m)_{k}}a^{(\ell)}_{k,m}=0 \quad
    \textnormal{for all} \quad 2\leq m \leq \ell+1
\end{equation}
Of course it is sufficient to prove
\begin{eqnarray}
  &&  \sum^{\ell}_{k=0}\frac{k!}{(m)_{k}}Q^{(\ell)}_{k,r}=0 ,\label{b7}
    \\
  &&  \textnormal{for all}\quad \{r,m |\, 0\leq r\leq \ell+1-m \, \quad 2\leq m \leq
  \ell+1\} .\nonumber
\end{eqnarray}
(\ref{b7}) can be verified easily for simple cases $m=\ell+1 \,,
r=0$\, or\, $m=\ell \,, r=0 \,\,\textnormal{or}\,\, r=1$. In general
we prove it by computer.

From (\ref{ms1}) and (\ref{b2}) we derive an integration mapping for
any fixed $m$ which acts on
$\left\{a^{(\ell)}_{k,m}\right\}^{\ell}_{k=0}$
\begin{equation}\label{b8}
    a^{(\ell)}_{k,m}\rightarrow \tilde{b}^{(\ell-1)}_{k,m-1}=
    \sum^{\ell-k}_{n=1}\frac{a^{(\ell)}_{k+n,m}}{(m+k-1)_{n}}P_{n-1}(k)
    ,
\end{equation}
By trying on a computer one can show that this mapping can be
repeated (with $(\ell, m)$ next replaced by $(\ell-1, m-1)$) exactly
$n=m-1$ times. The resulting coefficient is denoted
\begin{equation}\label{b9}
    b^{(\ell-n)}_{k,n} ,
\end{equation}
so that (\ref{multigrad1}) holds. It depends in fact on all three
parameters $\ell,k,n$. For $n=\ell$ it can be proved that
\begin{equation}\label{b10}
    b^{(0)}_{0,\ell}=1 ,
\end{equation}
whereas for $n=\ell-1$ we have guessed from a finite number of
examples
\begin{equation}\label{b11}
b^{(1)}_{0,\ell-1}=-\frac{1}{2}(\ell-1)_{4}\quad ,
b^{(1)}_{1,\ell-1}=-(\ell)_{2} .
\end{equation}

 A simple consequence
of (\ref{multigrad1}) and (\ref{b10}) is that the maximal singular
terms in $\Psi^{(\ell) mixed}$ are
\begin{equation}\label{b12}
\Psi^{(\ell) mixed}_{max.
sing}=[(a\cdot\nabla_{1})(c\cdot\nabla_{2})]^{\ell}(\zeta-1)^{-1}
\end{equation}

\section*{Appendix C}
\setcounter{equation}{0}
\renewcommand{\theequation}{C.\arabic{equation}}

We make an ansatz for $g^{p}_{k}$ $(p\geq 1 )$
\begin{eqnarray}
  && g^{p}_{k}=\sum^{[p/2]}_{n=0}\alpha_{p,n}\Delta_{p-n}
  (\ell+k-2(p-n-1))_{p-2n}+\frac{(-1)^{p}\gamma_{p}}{(\ell+k+2-p)_{p}}\label{c1}
\end{eqnarray}
and show that it is consistent with (\ref{rec}) if $\alpha_{p,n}$,
$\gamma_{p}$ satisfy the recursion relations
\begin{eqnarray}
  && \alpha_{p+1,n+1}=\frac{(p+1)(p-2n)}{2(n+1)}\alpha_{p,n} ,\label{c2} \\
  && \gamma_{p+1}=\frac{1}{2}p\gamma_{p} ,\label{c3}
\end{eqnarray}
which by the initial conditions
\begin{eqnarray}
  && \alpha_{p,0}=1   \quad\quad \textnormal{(by definition)}\nonumber \\
  && \gamma_{1}=\frac{1}{2}\label{c4}
\end{eqnarray}
imply
\begin{eqnarray}
  && \alpha_{p,n}=\frac{1}{2^{n}}n!\binom{p}{n}\binom{p-n}{n}, \label{c5}\\
  && \gamma_{p}=\frac{1}{2^{p}}(p-1)! .\label{c6}
\end{eqnarray}
 So we need to prove consistency of the ansatz (\ref{c1}) and the
 two recursion relations (\ref{c2}), (\ref{c3}).

 Eqn. (\ref{rec}) is already written in a form that is suited for
 our strategy. Inserting (\ref{c1}) in the r.h.s. of (\ref{rec}), we
 obtain a difference equation for $g^{p+1}_{k}$ of first order
 which is solved by summation. The result is compared with the
 ansatz. After insertion of (\ref{c1}) the r.h.s. of (\ref{rec}) is
\begin{eqnarray}
  && -(p+1)\left\{\sum^{[p/2]}_{n=0}\alpha_{p,n}(p-2n)
  \Delta_{p-n}[(\ell+k-2p)_{2n+3}]^{-1}\right. \nonumber\\
  && \left.-(-1)^{p}p\gamma_{p}[(\ell+k-2p)_{2p+3}]^{-1}\right\} .\label{c7}
\end{eqnarray}
For the summation we replace $k$ by $k'$ and sum
$\sum^{k-1}_{k'=0}$. In the first term of (\ref{c7}) we have
\begin{eqnarray}
  && \sum^{k-1}_{k'=0}[(\ell+k'-2p)_{2n+3}]=\frac{1}{2(n+1)}\left[\frac{(\ell-1-2p)!}
  {(\ell+1-2(p-n))!}\right.\nonumber \\
  && \left.-\frac{(\ell+k-1-2p)!}{(\ell+k+1-2(p-n))!}\right] .\label{c8}
\end{eqnarray}
In the second term of (\ref{c7}) we have an analogous expression to
sum with $n$ replaced by $p$.

On the l.h.s. of (\ref{rec}) we have after summation
\begin{equation}\label{c9}
    \frac{g^{p+1}_{k}}{(\ell+k-2p)_{p+1}}-\frac{g^{p+1}_{0}}{(\ell-2p)_{p+1}}
    .
\end{equation}
Since for $k=0$ there is no $p$ allowed by (\ref{rec}), we cancel
$g^{p+1}_{0}$ against the $k$- independent terms (resulting from
$k'=0$) on the r.h.s. Multiplying both sides with
\begin{equation}\label{c10}
    (\ell+k-2p)_{p+1} ,
\end{equation}
and using
\begin{equation}\label{c11}
    \frac{(\ell+k-2p)_{p+1}(\ell+k-1-2p)!}{(\ell+k+1-2(p-n))!}=(\ell+k-2(p-n-1))_{p-2n-1}
    ,
\end{equation}
the consistency of our ansatz and the correctness of (\ref{c2}),
(\ref{c3}) are easily inspected.

\end{document}